\documentclass[a4paper]{article}

\usepackage{INTERSPEECH_v2}
\usepackage{CJKutf8}
\title{Exploring Retraining-Free Speech Recognition\\for Intra-sentential Code-Switching}
\name{Zhen Huang, Xiaodan Zhuang, Daben Liu, Xiaoqiang Xiao, \\Yuchen Zhang, Sabato Marco Siniscalchi}
\address{Apple Inc., 1 Broadway, Cambridge, MA 02142}
\email{zhen\_huang@apple, xiaodan@apple.com, daben\_liu@apple.com, xiaoqiang\_xiao@apple.com, yuchen\_zhang@apple.com, ssiniscalchi@apple.com}

\begin{document}
 \ninept
\maketitle
\begin{abstract}
In this paper, we present our initial efforts for building a code-switching (CS) speech recognition system leveraging existing acoustic models (AMs) and language models (LMs), i.e., no  training required, and specifically targeting intra-sentential switching. To achieve such an ambitious goal, new mechanisms for foreign pronunciation generation and language model (LM) enrichment have been devised. Specifically, we have designed an automatic approach to obtain high quality pronunciation of foreign language (FL) words in the native language (NL) phoneme set using existing acoustic phone decoders and an LSTM-based grapheme-to-phoneme  (G2P) model. Improved accented pronunciations have thus been obtained by learning foreign pronunciations directly from data. Furthermore, a code-switching LM was deployed by converting the original NL LM into a CS LM  using translated word pairs and borrowing  statistics for the NL LM.  Experimental evidence clearly demonstrates that our approach better deals with  accented foreign pronunciations  than techniques based on human labeling. Moreover, our best system achieves a 55.5\% relative word error rate reduction from 34.4\%, obtained with a conventional monolingual ASR system,  to 15.3\% on an intra-sentential CS task without harming the  monolingual recognition accuracy.

\end{abstract}

\section{Introduction}
Code-switching (CS) is usually referred to as the situation where a speaker alternates between different languages within a single conversation, e.g.,  \cite{poplack_1980, sankoff_1981, milroy_1995, auer2013code}.  Code-switching can be broadly divided into two groups  \cite{myers_scotton_1993}: inter-sentential switching -  the alternation is between sentences (also called extra-sentential),  and intra-sentential  switching - the alternation is within sentences (it can also include intra-word).  With the rapidly growing of bilingual/multilingual population,  CS is no longer a  phenomenon relevant for  minority languages, which are affected by  majority languages, but it also concerns majority languages influenced by  \emph{lingua francas}, such as English and French, as properly pointed out in \cite{yilmaz_asru17}.     Despite the recent significant advances witnessed in the field of  automatic speech recognition (ASR) \cite{Hinton_2012}, ASR systems have unfortunately still limited capability in tackling  the code-switching problem, especially intra-sentential switching. Major challenges for CS speech recognition include different phone sets among languages, and insufficient intra-sentential code-switching training data.  

In recent years, we have witnessed an increasing research effort to address the CS problem, and available approaches  can  mainly be divided  into two groups: robust acoustic modeling,  and language modeling.  For example,  a global phoneme set is constructed by merging phoneme sets from different languages, and a unified multilingual acoustic model (AM) is trained in  \cite{yu2004chinese, huang2007generation, lin2009study}. However, the goal of those studies on cross-language AMs is to develop ASR systems for new languages for which scarce or no training material is  available. In the CS scenario,  the native language (NL) AM should be used at the runtime, and the recognition accuracy and speed for the NL ASR should not  be sacrificed.  In \cite{yu2009cross}, the authors compared four ASR approaches to tackle the CS task under the constraint that the NL AM is to be used at runtime. In those approaches, the lexicon in the foreign languages (FLs)  are represented by the NL phoneme set through phone/senone mapping/merging using either knowledge-based or data-driven mapping techniques, and the NL acoustic model (AM) is employed to carry out recognition. It was shown that top recognition results were obtained by mapping each senone in the FL's AM to the senone in the NL's AM with the minimum Kullback-Leibler divergence. From a language model (LM) point of view,   a statistical machine translation (SMT) based text generation technique  was used in  \cite{vu2012first} to artificially augment the amount of training data to estimate more robust LMs. In \cite{weng1997study},  a multi-lingual  LM is trained by combining monolingual LMs into a probabilistic finite state grammar. More recently, syntactic and semantic features were integrated into a factored language model to combat data scarcity in code-switching speech \cite{adel2015syntactic}.  Recurrent neural networks (RNNs) and factored language models are instead combined in \cite{adel2014combining} for improved CS performance. In \cite{ma2015recognize}, the authors successfully  used similar word pairs to borrow information from available (already trained) monolingual LMs and boost foreign low-frequency words.  Since code-switching deals with more than a language, it is natural to exploite language identification, e.g., \cite{waibel2000multilingual, lin2012recognition, gonzalez2015real}, or  language diarization  \cite{yilmaz_asru17} to select among monolingual speech recognizers. Language identification was also employed in \cite{vu2012first} to be combined with AM scores for better recognition performance. However, those approaches suffers from errors propagated by the language identification/diarization module to the ASR block.

In the afore-mentioned approaches, either ad-hoc acoustic and/or language models have to be built to specifically handle the CS task, or  multiple language-dependent recognizers preceded by a language detection step have to be employed.  In this paper, we demonstrate that leveraging existing monolingual ASR resources is possible to achieve effective intra-sentential code-switching without retraining any of the ASR models, namely  the acoustic and language model, without sacrificing the recognition accuracy of the NL ASR system. We attain such a challenging goal by devising new mechanisms for foreign pronunciation generation, and LM enrichment. Specifically, we generate a high quality foreign pronunciation lexicon by obtaining the pronunciation of FL words in the NL phoneme set through acoustic phone decoding, and grapheme-to-phoneme conversion (G2P). Differently  from \cite{yu2009cross}, where  phone/senone mapping and merging was the focus, we adopt data-driven techniques that learn foreign pronunciations directly from data, which in turn allows us to better deal with accented pronunciations. The experimental evidence interestingly demonstrates that data-driven acoustic phone decoding and G2P approaches perform favorably with respect to manual labeling. Moreover, we have obtained an enriched CS LM  taking inspiration from  \cite{ma2015recognize}. Specifically, we have converted the initial NL LM into a CS LM by  borrowing statistics/information from translated word pairs, which allows us to avoid LM retraining. Comparing to the state-of-the-art approach in \cite{vu2012first}, the proposed method doesn't need to construct a new phone set and retrain the AM, and also saves the efforts to retrain the LM. The retained LM by collecting/generating new CS data as in \cite{vu2012first} can easily harm the recognition performance on NL.  Experimental evidence on introducing FL words to a NL system with vocabulary size in the hundreds of thousands demonstrate that our best system is able to provide 55.5\% relative word error rate reduction (from 34.4\% to 15.3\%) on a code-switching testing set without harming the performance on general NL test sets. 

%The rest of this paper is organized as follows. Section \ref{sec:PG} will discuss the methods used to obtain FL pronunciations.  In Section \ref{sec:LM}, we describe the CS LM construction by enriching the NL LM. We report and analyze  the experimental results in Section \ref{sec:exp}, followed by discussion and future work in  Section \ref{sec:discussion}. 

\section{Foreign Pronunciation Generation}
\label{sec:PG}
NL refers the language whose phoneme set and AM are used in recognition. All other languages are indicated as FLs. For each FL word  in a CS utterances, a native pronunciation representation in the NL's phoneme set is generated.

\subsection{Linguist Manual Labeling}
\label{ssec:ll}
Manual labeling by  experts can be the most straightforward way for this kind of pronunciation generation task. Linguists with knowledge of the NL and a FL can be hired to directly produce the lexicon of FL words with NL's phoneme set. The strength of this approach is  pronunciations could be always reasonable for human perception; the shortcoming could be that linguists tend to label the FL words with the standard pronunciation, but native people always has some extent of accent when they speak foreign words  in a CS scenario.  

\subsection{Phone Decoding}

Phone decoding is a data-driven method to obtain pronunciations for the FL words. When the audio segments containing FL words, a NL ASR system can be used to decode those segments and thus obtain the possible phoneme sequences of the FL words. The audio segments mainly come from two sources: foreign people speaking FL words, and native people speaking FL words. 

\subsubsection{Phoneme Confusion Network}
\begin{figure}[h]
	\centering
	%\centerline{\includegraphics[width=0.8\linewidth]{nn}}
	\includegraphics[width=0.7\linewidth]{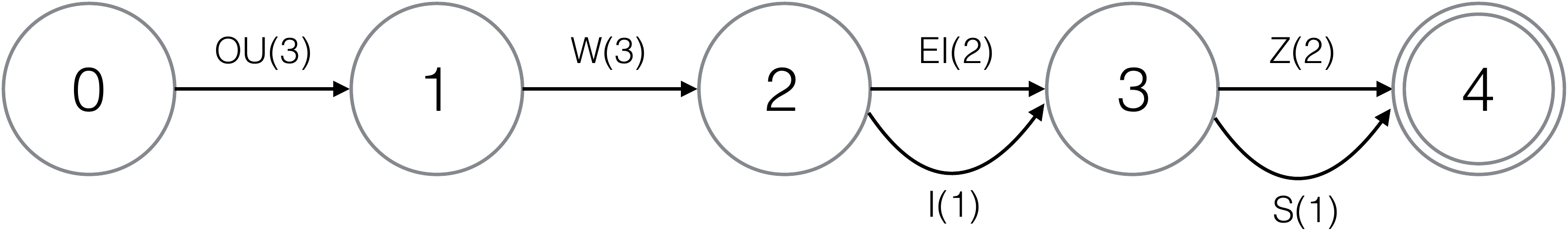}
	\caption{Phoneme confusion network for the word ``always" }
	\label{fig:cn}
	\vspace{-3mm}
\end{figure}

In harvesting audio segments for FL words, a particular word usually corresponds to more than one audio segment. The decoded phoneme sequences for those audio segments may vary a lot, and the process to obtain a good phoneme representation by summarizing all decoded phoneme sequences becomes crucial. In this paper, we adopt a ROVER-like \cite{fiscus1997post}  method based on  phoneme confusion networks (CNs) to accomplish the summarization process. For example, let's suppose that there exist three (3) audio segments containing the  English word ``always," and the phone sequences generated with a Chinese ASR system are:
\begin{itemize}
	\item OU W EI Z
	\item OU W I Z
	\item OU W EI S
\end{itemize}
Then the corresponding phoneme confusion network is displayed in  Figure \ref{fig:cn}, where the numbers in  parentheses represent the number of times a phoneme appears. From the phoneme confusion network in Figure \ref{fig:cn}, we can conclude that the best phoneme sequence (with most votes) for ``always"  is ``OU W EI Z," and the second best could either be ``OU W EI S," or ``OU W I Z." For each FL words, we can retain an N-best representation.

\subsubsection{\textbf{P}hone \textbf{D}ecoding on \textbf{F}L words spoken by \textbf{F}oreign people (\textbf{PDFF})}
\label{sssec:pdf}
In this paper, we suppose we have an already-trained monolingual speech recognition system for each language and there are plenty of corresponding training utterances. To harvest audio segments for foreign words, we use the FL speech recognition system to performance force alignment on its training utterances, by doing so, we can obtain the start/end time stamps of each word in those utterance. Then the audio segments containing concerned words are decoded with NL speech recognition system generating possible phoneme sequences for those words. 

The advantage of this approach is that large amount of training data can be utilized to generate a lot of possible pronunciation representations and summaries them to get the best guesses, but a major concern is that those foreign words are spoken by foreign people, so to some extend, we are actually obtaining ``standard" pronunciations while what we really care about is native people speaking foreign words with possible accents. To obtain such pronunciation guessing, we try to make use of the NL training data.

\subsubsection{\textbf{P}hone \textbf{D}ecoding on \textbf{F}L words spoken by \textbf{N}ative people (\textbf{PDFN})}
\label{sssec:pdn}
Differently from FL training data, there are much fewer examples in NL training data that contain foreign words. Since we don't have native pronunciation in the native phoneme set for those words, it is not possible  to obtain forced-alignments and find the corresponding audio segments (for those words) as shown in Section \ref{sssec:pdf}. However,  foreign words appearing in the NL training data are  the actual CS words, and  although  limited amount,  these words can still be fundamental  to find more relative pronunciation guesses for CS speech recognition. To obtain the audio segments of those foreign words in NL training datasets, we first utilize the PDFF system (see  Section \ref{sssec:pdf}) to obtain an initial lexicon, and we then perform forced-alignment with such an initial lexicon to find the location of those words. After finding those audio segments, we can follow Section \ref{sssec:pdf} to generate the pronunciation guesses.

\subsection{Grapheme-to-Phoneme Conversion}
\label{ssec:G2P}
Grapheme-to-Phoneme conversion(G2P) is the process of converting the written form of a word to its pronunciation. In this paper, we use a Long Short-Term Memory (LSTM) \cite{hochreiter1997long} based G2P model to translate the FL words from written form to pronunciation representation in NL's phoneme set. Our G2P model adopts an encoder-decoder architecture with attention mechanism \cite{bahdanau2014neural} which is different from the LSTM G2P model proposed in \cite{rao2015grapheme}. There are two sources of data we used for the G2P model training: 
\begin{itemize}
\item[(a)]  Standard NL lexicon containing linguist pronunciation labeling of both NL words and FL words. The amount of FL words is limited in this scenario;
\item[(b)]  Phone decoding results on FL words spoken by native people (i.e., PDFN in Section \ref{sssec:pdn})
\end{itemize}
The training data (b) is added for the purpose of capturing the accents of NL speakers when they speak FL words. To obtain training data (b), we went through the following steps:
\begin{itemize}
	\item[1] Collect audio recordings for the 5000 most popular FL words, which are spoken by NL speakers. 
	\item[2] Use G2P trained with data (a) to generate 4 pronunciations from written form for each FL word used in Step1.
	\item[3] Use a NL phone decoder to generate additional 4 pronunciations for each FL word provided the audio collected in Step1.   
	\item[4] The combined 8 pronunciations from Step2 and Step3 are rescored using the NL acoustic model, and the top 4 pronunciations with the highest acoustic model scores are retained for each FL word. 
\end{itemize}
Then (a) and (b) are combined to form the final training data for the LSTM G2P model training. 

\section{Language Model Enriching}
\label{sec:LM}
With the FL word pronunciation problem tackled, we still have to fix the LM part to deploy a robust  CS system. LM training is  a crucial problem in CS speech recognition, and the chief difficulty is in finding enough intra-sentential CS  training data to build a reliable  CS LM. In fact, the CS phenomenon does not happen as frequently in  written form. However,  it should not be difficult to convince ourselves that the lack of abundant CS text material is not a real concern, since the performance of the  NL ASR system would be harmed if too much CS data is used to train the LM.  Therefore,  we started seeking into possible ways to overcome improve the LM in a CS system without collecting ad-hoc CS LM data, and we  the  idea proposed in \cite{ma2015recognize} to be a sound solution to construct a CS LM. The idea is to borrow statistics/information in an already-trained LM, and we adopt and extend this idea in the present work. 

In code-switching,  we usually replaces a native word with the same meaning word (translated counterpart) in foreign language. For example, the monolingual Chinese utterance (meaning in English: ``We play basketball")
\begin{itemize}
	\item \begin{CJK}{UTF8}{gbsn}		
		我们 打 篮球	
	\end{CJK}
\end{itemize}
can be spoken as
\begin{itemize}
	\item \begin{CJK}{UTF8}{gbsn}		
		我们 打 basketball
	\end{CJK}
\end{itemize}
in code-switching scenarios. The code-switched words are a translated word pair ``\begin{CJK}{UTF8}{gbsn}篮球\end{CJK}--basketball". Here ``basketball" has the same meaning with ``\begin{CJK}{UTF8}{gbsn}	篮球	\end{CJK}". In this example, when we construct the CS LM, we add the English (FL) word ``basketball" into Chinese (NL) LM by copying the statistics/information of its translated counterpart ``\begin{CJK}{UTF8}{gbsn}	篮球	\end{CJK}". By doing so, we borrow statistics/information of the word ``\begin{CJK}{UTF8}{gbsn}篮球\end{CJK}" in the NL LM for the added FL word ``basketball".  The effort of collecting/generating CS data and retraining LM can thereby be avoided using such an elegant  statistics-borrowing method.

\begin{figure}[t]
	\centering
	%\centerline{\includegraphics[width=0.8\linewidth]{nn}}
	\includegraphics[width=0.7\linewidth]{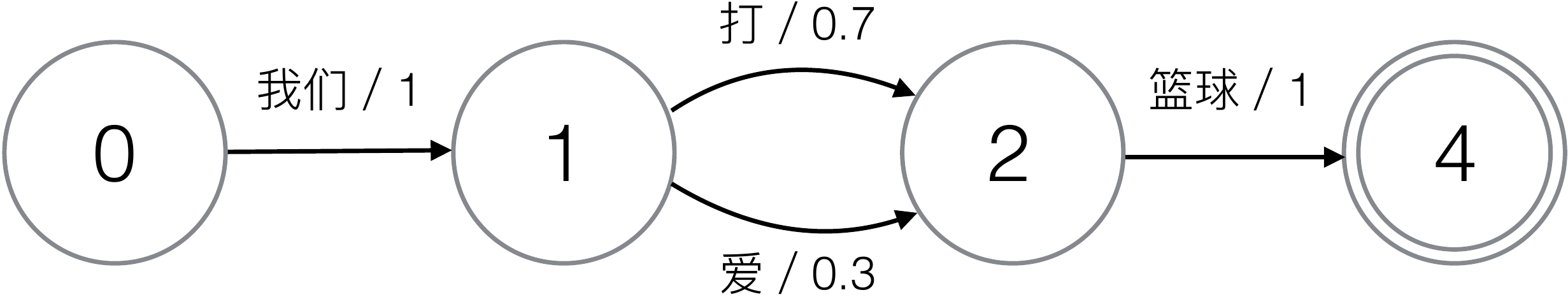}
	\caption{NL LM G}
	\label{fig:fst1}
	 \vspace{-4mm}
\end{figure}

\begin{figure}[t]
	\centering
	%\centerline{\includegraphics[width=0.8\linewidth]{nn}}
	\includegraphics[width=0.7\linewidth]{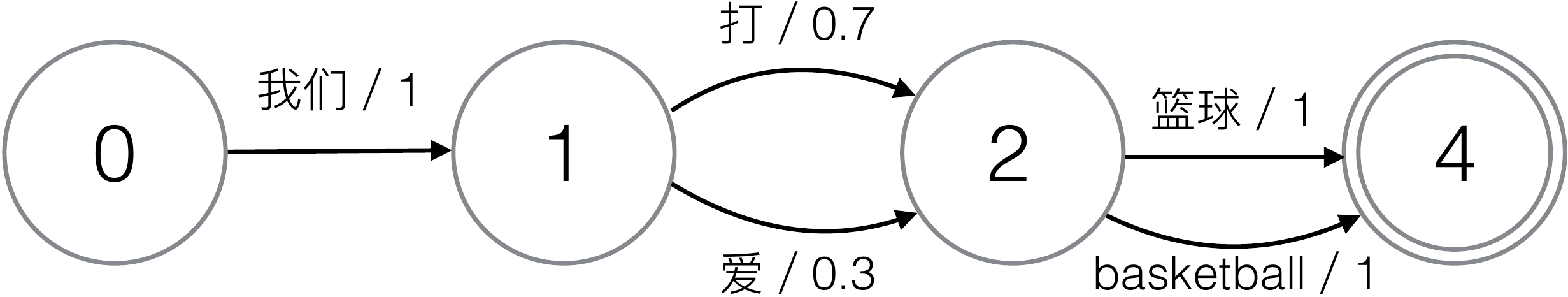}
	\caption{CS LM G'  converted from NL LM G in Figure \ref{fig:fst1} by adding FL word ``basketball"}
	\label{fig:fst2}
			  \vspace{-4mm}
\end{figure}

Modern ASR systems usually use weighted finite-state transducers (WFST) \cite{mohri2002weighted} to present the LM as well as  other components. %WFSTs  provide a common and natural representation for components of ASR system and general transducer operations combine these representations flexibly and efficiently \cite{mohri2002weighted}, also some common decoding operations such as pruning are more manageable under the framework of WFST \cite{Paulik2015ImprovementsTT}. 
We  therefore show how to  convert an NL LM into a CS LM in the context of WFST.  In the LM WFST graph G, each transition/edge  corresponds to a word, and the weight of the edge indicates the word's probability. Figure \ref{fig:fst1} is a toy example of a native LM's G graph. When we convert this toy NL LM into a CS LM concerning only the word ``basketball" as follows: (i) Obtain the Chinese counterpart word ``\begin{CJK}{UTF8}{gbsn}篮球\end{CJK}" of ``basketball,"  (ii) Find the edge in the WFST corresponding to the word ``\begin{CJK}{UTF8}{gbsn}篮球\end{CJK}," and (iii) Insert a parallel edge with the same weight into the WFST but carrying the the foreign word  ``basketbal.'' The above steps will result in the CS G graph shown in Figure \ref{fig:fst2}.


\begin{table}[t]\footnotesize
	\begin{center}
		\caption{Lexicons used in the experimental evaluation}
		  \vspace{-2mm}
		\label{exp:lex}
		\centerline{
			\begin{tabular}{|c||c|}
				\hline
				{\bf Lexicons} & {\bf Configurations and Methods}  \\
				\hline
				L0 &  original NL lexicon (no FL words) \\
				\hline
				L1 &  224 FL words by linguist labeling in Section \ref{ssec:ll}\\
				\hline
				L2f  &  1155 FL words by PDFF in Section \ref{sssec:pdf} \\
				\hline
				L2n &  224 FL words by PDFN in Section \ref{sssec:pdn}  \\
				\hline
				L3a &  1155 FL words by G2P with data (a) in Section \ref{ssec:G2P}\\
				\hline
				L3ab & 1155 FL  words by G2P with data (a) and (b) in Section \ref{ssec:G2P}\\
				\hline		
		\end{tabular}}
	  \vspace{-7mm}
	\end{center}
\end{table}	

\begin{table}[t]\footnotesize
	\begin{center}
		\caption{LMs  with different FL vocabularies (see Section \ref{sec:LM})}
		  \vspace{-2mm}
		\label{exp:lm}
		\centerline{
			\begin{tabular}{|c||c|}
				\hline
				{\bf LMs} & {\bf Configurations}  \\
				\hline
				G0 &  original NL LM (no FL words) \\
				\hline
				G1 &  CS LM with 224 FL words\\
				\hline
				G2  &  CS LM with 1155 FL words \\
				\hline	
		\end{tabular}}
	  \vspace{-7mm}
	\end{center}
\end{table}

%If we add more words in the CS LM, for example, ``We", ``play" and ``love" (translated counterpart of ``\begin{CJK}{UTF8}{gbsn}我们\end{CJK}", ``\begin{CJK}{UTF8}{gbsn}打\end{CJK}" and ``\begin{CJK}{UTF8}{gbsn}爱\end{CJK}", respectively), the CS G graph will become Figure \ref{fig:fst3}. 

%\begin{figure}[h]
%	\centering
%	\includegraphics[width=\linewidth]{figs/FST3-crop}
%	\caption{CS language model G  graph concerning more words}
%	\label{fig:fst3}
%\end{figure}

%Generally speaking, when we add a FL word ``A" into NL LM's G graph, we obtain the FL word's translated counterpart word ``B" in NL, add a parallel edge for every corresponding edge of word ``B" and copy the weight of original transition. 

\section{Experimental Setup \& Results}
\label{sec:exp}
 Our approach is validated using  Chinese as the NL, and English as the FL.  
\vspace{-2mm}
\subsection{Experimental Setup}
\label{sec:setup}
{\bf Acoustic Model}: The NL Chinese acoustic model is a Feed-Forward deep neural network (FFDNN) trained with filter bank features from 160 hours Chinese speech using cross-entropy and bMMI objective function. The FL English acoustic model utilize similar setup.
%The NL Chinese acoustic model is a Feed-Forward deep neural network (FFDNN). The DNN is constructed with 6 hidden sigmoid layers and a output softmax layer.  Each hidden layer has 1024 neurons.  The DNN is trained with filter bank features from 160 hours Chinese speech. It is initialized with stacked restricted Boltzmann machines by using layer by layer generative pre-training \cite{hinton2006reducing}, and then trained with cross-entropy and bMMI objective function. The FL English acoustic model is also a FFDNN trained with 160 hour of English speech. 
%\subsection{Lexicons}
%\label{ssec:lexicon}

\noindent{\bf Lexicon}: Table \ref{exp:lex} is the lexicons we obtained by methods proposed in Section \ref{sec:PG}. The 1155 FL words are selected by their appearing frequencies in the English training data and also the probability occurring during Chinese-English code-switching. The 224 FL words is a subset of the 1155 words which actually occur in the Chinese training data.

\noindent{\bf Code-Switching LM}:  Table \ref{exp:lm} shows the code-switching LMs obtained following Section \ref{sec:LM}. We refer to the original Chinese LM as G0, the CS LM contains the 224 English word as G1, and the CS LM contains all the 1155 words as G2.

\noindent{\bf Testing sets}: There are two test sets. The code-switching test set contains 107 CS utterances. Each of the utterances contains 1 or more intra-sentential code-switchings. A typical one is like ``	\begin{CJK}{UTF8}{gbsn}	我们 打 basketball\end{CJK}". The FL words included in the CS test set is mainly a subset of the 224 English words afore-mentioend. The general test set contains 9677 utterances, most of which are pure Chinese utterances.

\noindent{\bf Evaluation Metric}: Experimental results are reported in terms of a slightly modified version of the conventional word error rate (WER) metric, namely each Chinese character is treated as a word in the contest of CS speech recognition.

\begin{table}[t]\footnotesize
	\begin{center}
		\caption{WERs (in \%) on CS test set by using different lexicon and LM configurations. Word error rates reductions (WERRs)  are given in parentheses.}
		  \vspace{-2mm}
		\label{exp:CS}
		\centerline{
			\begin{tabular}{|c||c||c|}
				\hline
				1&{\bf Configuration} & {\bf WER on CS}  \\
				\hline
				2&L0+G0 (baseline) &  34.4\%  \\
				\hline
				4&(L0+L2f)+G1 &  24.3\%  (29.4\%)\\
				\hline
				3&(L0+L1)+G1 &  24.3\% (29.4\%)\\
				\hline
				5&(L0+L2n)+G1 &  20.7\%  (39.9\%)\\
				\hline
				6&(L0+L3a)+G1 &  21.6\% (37.2\%) \\
				\hline
				7&(L0+L3ab)+G1 &  18.8\% (45.3\%) \\
				\hline
				8&(L0+L1+L2f+L2n+L3ab)+G1 &  15.3\% (55.5\%)  \\
				\hline
				9&(L0+L1+L2f+L2n+L3ab)+G2 &  15.3\%  (55.5\%) \\
				\hline
		\end{tabular}}
	\vspace{-7mm}
	\end{center}
\end{table}	

\begin{table}[th]\footnotesize
	\begin{center}
		\caption{WERs (in \%) on the  general test set.}
				  \vspace{-2mm}
		\label{exp:general}
		\centerline{
			\begin{tabular}{|c||c|}
				\hline
				{\bf Configuration} & {\bf WER on general}  \\
				\hline
				L0+G0 (baseline) &  14.5\%  \\
				\hline
				(L0+L1+L2f+L2n+L3ab)+G1 &  14.5\%  \\
				\hline
				(L0+L1+L2f+L2n+L3ab)+G2 &  14.6\%  \\
				\hline
		\end{tabular}}
	  \vspace{-7mm}
	\end{center}
\end{table}

\subsection{Experimental Results}
Table \ref{exp:CS} shows the CS performance of several systems having different lexicon and LM configurations. The CS system employing the original NL lexicon, namely L0, and the original LM, namely G0, is referred to as the baseline system. By comparing the second and third row in in Table \ref{exp:CS},  we can easily argue that the proposed PDFF for generating the lexicon of FL words (see Section \ref{sssec:pdf}) along with enriched LM using 224 FL words (G1) delivers significant performance improvement over the baseline system. Indeed,  the WER is reduced from 34.4\% down to 24.3\%, which is equivalent to a  word error rate reduction (WERR) of 29.4\%. The monolingual Chinese ASR system (baseline) cannot conversely properly recognize any foreign words, which are treated as out of vocabulary words. Moreover, we can observed that the proposed PDFF approach is equivalent,  in terms of recognition performance, to the manual labeling of an expert (compare  third and fourth rows). The latter might allow us to conclude that the data-driven method - without taking into account speaker's accents yet,  is comparable to an expensive human-driven solution. By taking into account the speaker's accent using the L2n lexicon, we can further reduce the WER down to 20.7\%, which represents a WERR of 39.9\%, outperforming  both human labeling, and the PDFN approach that tries to only  generate ``standard" pronunciations. 

If an LSTM-based G2P approach trained on both the standard NL lexicon covering manually labeled NL and FL words (i.e., set (a) in Section \ref{ssec:G2P}), and the phone decoding results (i.e., set (b) in Section \ref{ssec:G2P}), a WER of 18.8\% is attained, as displayed in the seventh row.  By comparing the latter result with that shown in the sixth row in Table \ref{exp:CS}, which corresponds to G2P train on the set (a) only, we can see that taking speaker's accent into account has a  beneficial effect on the recognition performance. Merging together all the pronunciation guesses from different sources brings the  WER is down to 15.3\%, as shown in the eighth row in Table \ref{exp:CS}, which accounts for a WERR of 55.5\%. The latter outcome seems to imply that those pronunciation generation methods are complementary to each other.  Finally, we noticed that increasing the amount of FL words in the proposed CS LM does not provide any further improvement, but it does not  harm the recognition performance on CS test set either.  It is also important to  verify that the addition of ad-hoc CS lexicons and LMs does not cause a drop in baseline ASR performance on pure Chinese utterances.  To this end, we ran a new set of experiments on a more general data set (see Section \ref{sec:setup}). The corresponding results are reported the results in  Table \ref{exp:general}, and these results demonstrate that the  performance degradation on a more general ASR task is ignorable. 

Finally,  the effect of tuning the weight-copying scale in LM enriching is given in Table \ref{exp:tuning}. A scale factor  is multiplied  by the original NL edge's weight before being copied in the added parallel FL edge. A value greater than 1 implies a boost of the FL edge, whereas a value smaller than 1  suppresses the FL edge. From Table \ref{exp:tuning}, we can see that boosting FL edges improves the recognition performance on the CS test, but it also causes a degradation on the general test set. Suppressing FL edges won't harm the performance on the general set but causes a degradation on the CS set.   That can easily be understood by bearing in mind that  errors on the CS test set are mainly caused by  unrecognized FL words; therefore, better results can be obtained by  boosting the FL edges. However, there are nearly no FL words in a more general test set, and that explains the drop in performance. 

 \vspace{-3mm}
\begin{table}[h]\footnotesize
	\begin{center}
		\caption{Effects on tuning the weight-copying scale with configuration of (L0+L1+L2f+L2n+L3ab)+G2. }
		\vspace{-2mm}
		\label{exp:tuning}
		\centerline{
			\begin{tabular}{|c||c|c|}
				\hline
				{\bf Scale} & {\bf WER on general}& \bf WER on CS \\
				\hline
				1 &  14.5\% &  15.3\%\\
				\hline
				1.5 &  14.9\% &  12.8\%\\
				\hline
				0.667 &  14.5\% & 18.6\%\\
				\hline
		\end{tabular}}
	  \vspace{-7mm}
	\end{center}
\end{table}	

\section{Conclusion}
\label{sec:discussion}
In this work, we build a intra-sentential code-switching ASR prototype  by leveraging existing monolingual systems without either acoustic, or language model retraining. Our approach focuses on data-driven pronunciation generation for FL words and takes into account speakers' accent. Moreover, we also devise  a simply  CS LM generation approach by enriching existing monolingual LM leveraging word pairs. The significant accuracy improvement on CS test data together with retention of LM accuracy on a more general speech task highlights the effectiveness of our idea. 
A major limitation of the proposed approach  concerns the  LM enriching method, which may not handle the word reordering problem. To clear ideas, let's consider the English word ``tomorrow"  that corresponds to the  Chinese word ``\begin{CJK}{UTF8}{gbsn}明天\end{CJK}". In  ``I will go to school tomorrow"  the adverb ``tomorrow" appears at the end of the sentence, but the same sentence in Chinese, namely   ``\begin{CJK}{UTF8}{gbsn}我 明天 去 学校\end{CJK}“'' would  places  ``\begin{CJK}{UTF8}{gbsn}明天\end{CJK}" in the middle of the sentence. In a CS sentence having   ``tomorrow'' as the FL word,  `tomorrow''  would rarely appear in the middle of the sentence, as it should in  Chinese, and that may make the use of the  statistics borrowed from the NL LM for the CS LM not accurate. We will try to tackle this issue by combing the proposed statistic-borrowing and the traditional data augmentation methods.
\newpage
%%There are also other directions to polish the proposed system. For CS LM construction, we will explore leveraging statistics from a FL LM as well neighboring words in the NL text, together with NL LM statistics, for better robustness. Besides, such statistics could allow effective word-specific scaling for weight copying.
%%We will also enlarging the CS test set for more comprehensive evaluation. 
%%
%%Note that this is only a prototype system which is not incorporated into Apple Siri due to practical concerns. 
%%\section{Acknowledgments}
%%The authors would like to thank the team lead by Matthias Paulik at Apple Inc. to provide the LSTM G2P tool and Chao Weng, Bing Zhang, Rongqing Huang at Apple Inc. for valuable discussions and helps .
%\newpage
\bibliographystyle{IEEEtran}
\bibliography{mybib}
\end{document}